\documentclass[sigconf, screen]{acmart}

\AtBeginDocument{%
  }

\setcopyright{acmlicensed}
\copyrightyear{2018}
\acmYear{2018}
\acmDOI{XXXXXXX.XXXXXXX}
\acmConference[Conference acronym 'XX]{Make sure to enter the correct
  conference title from your rights confirmation emai}{June 03--05,
  2018}{Woodstock, NY}

\acmISBN{978-1-4503-XXXX-X/18/06}

\copyrightyear{2025}
\acmYear{2025}
\setcopyright{cc}
\setcctype{by-nd}
\acmConference[AutomotiveUI '25]{17th International Conference on Automotive User Interfaces and Interactive Vehicular Applications}{September 21--25, 2025}{Brisbane, QLD, Australia}
\acmBooktitle{17th International Conference on Automotive User Interfaces and Interactive Vehicular Applications (AutomotiveUI '25), September 21--25, 2025, Brisbane, QLD, Australia}
\acmDOI{10.1145/3744333.3747826}
\acmISBN{979-8-4007-2013-0/2025/09}

\begin{document}

\title[Uncertainty on Display]{Uncertainty on Display: The Effects of Communicating Confidence Cues in Autonomous Vehicle-Pedestrian Interactions}

\author{Yue Luo}
\email{leah00ly@163.com}
\orcid{}
\affiliation{Design Lab, Sydney School of Architecture, Design and Planning,
  \institution{The University of Sydney}
  \city{Sydney}
  \state{NSW}
  \country{Australia}
}

\author{Xinyan Yu}
\email{xinyan.yu@sydney.edu.au}
\orcid{0000-0001-8299-3381}
\affiliation{Design Lab, Sydney School of Architecture, Design and Planning
  \institution{The University of Sydney}
  \city{Sydney}
  \state{NSW}
  \country{Australia}
}

\author{Tram Thi Minh Tran}
\email{tram.tran@sydney.edu.au}
\orcid{0000-0002-4958-2465}
\affiliation{Design Lab, Sydney School of Architecture, Design and Planning,
  \institution{The University of Sydney}
  \city{Sydney}
  \state{NSW}
  \country{Australia}
}

\author{Marius Hoggenmueller}
\email{marius.hoggenmueller@sydney.edu.au}
\orcid{0000-0002-8893-5729}
\affiliation{Design Lab, Sydney School of Architecture, Design and Planning
  \institution{The University of Sydney} 
  \city{Sydney}
  \state{NSW}
  \country{Australia}
}

\renewcommand{\shortauthors}{Luo et al.}



\begin{abstract} %149 words
Uncertainty is an inherent aspect of autonomous vehicle (AV) decision-making, yet it is rarely communicated to pedestrians, which hinders transparency.  This study investigates how AV uncertainty can be conveyed through two approaches: explicit communication (confidence percentage displays) and implicit communication (vehicle motion cues), across different confidence levels (high and low). Through a within-subject VR experiment (N=26), we evaluated these approaches in a crossing scenario, assessing interface qualities~(visibility and intuitiveness), how well the information conveyed the vehicle’s level of confidence, and their impact on participants' perceived safety, trust, and user experience. Our results show that explicit communication is more effective and preferred for conveying uncertainty, enhancing safety, trust, and user experience. Conversely, implicit communication introduces ambiguity, especially when AV confidence is low. This research provides empirical insights into how uncertainty communication shapes pedestrian interpretation of AV behaviour and offer design guidance for external interfaces that integrate uncertainty as a communicative element.

%This research advances the understanding of how uncertainty communication influences pedestrians and provides valuable guidance for designing future eHMIs to effectively communicate uncertainty.
\end{abstract}

\begin{CCSXML}
<ccs2012>
   <concept>
       <concept_id>10003120.10003121.10011748</concept_id>
       <concept_desc>Human-centered computing~Empirical studies in HCI</concept_desc>
       <concept_significance>500</concept_significance>
       </concept>
 </ccs2012>
\end{CCSXML}

\ccsdesc[500]{Human-centered computing~Empirical studies in HCI}

\keywords{autonomous vehicles, uncertainty, AV-pedestrian interaction, human-machine interfaces, eHMIs}

% The term uncertainty is vaguely presented in the Introduction and Related Work. It requires more explanation about the context in which the uncertainty is being discussed. For example, is the uncertainty about speed or distance from the pedestrian? Provide more context early in the paper to make it easier to comprehend.

%The implicit or movement-based approach is abruptly presented in the introduction. More information is required to understand the implicit or movement-based approach to uncertainty communication.

\maketitle 
\section{Introduction}
In automated systems, uncertainty is an inherent characteristic that stems from factors such as incomplete or noisy data inputs and limitations in algorithmic design, leading to potential inaccuracies in decision-making outcomes~\cite{Bhatt2021}. This inherent uncertainty can, in turn, trigger unexpected or erratic behaviours, a particularly pressing issue in safety-critical systems such as autonomous vehicles (AVs), where even minor errors or failures can result in severe consequences~\cite{kunze2019}. Given the growing integration of AVs into public environments, effectively conveying system information to human actors, especially vulnerable road users like pedestrians, is crucial for ensuring safe interactions.

Effective communication of uncertainty serves as a key aspect of algorithmic transparency, playing a significant role in appropriately calibrating trust in automated systems~\cite{Prabhudesai2023}. 
%Despite its importance, uncertainty communication in automated decision-making remains underexplored~\cite{schum2014}, especially in the context of AV-pedestrian interactions, where uncertainty is often hidden from pedestrians. Although an increasing number of studies have investigated the  approaches and effects of conveying uncertainty to passengers or operators of AVs
In the context of AVs, uncertainty communication in automated decision-making has largely focused on in-vehicle contexts~\cite{schum2014}, with numerous studies examining how to visualise the AV's internal processes~(e.g., its detection and prediction capabilities) and the resulting impacts on passengers and drivers~\cite{Colley2021SemanticSegmentation,Helldin2013AVUncertainty,Colley2024Uncertain,kunze2019,Peintner2022, Kunze2018, Kunze2018b}. 
However, AV's uncertainty information, such as its confidence identifying pedestrian crossing intent and deciding whether to yield at unsignalised crossings, is largely concealed from pedestrians. This lack of transparency may lead to overtrust in AV systems due to a lack of awareness of their fallibility~\cite{kunze2019}, potentially resulting in serious safety risks for vulnerable road users.

Unlike in-vehicle users, pedestrians lack direct access to AV system information, making external human-machine interfaces (eHMIs)~\cite{Mahadevan2018, Locken2019, dey2020taming} and vehicle kinematics~\cite{Dey2017, Moore2019, chen2020comparison} critical channels for conveying AV's internal processes. While eHMIs have been widely studied for signalling AV's status and intent~\cite{dey2020taming}, their role in uncertainty communication remains underexplored. Research also shows that basic eHMI messages can aggravate effects of overtrust, particularly in erroneous AV operation~\cite{faas2021calibrating} or malfunctioning eHMI displays~\cite{hollaender2019}. Meanwhile, the vehicle’s movement plays a critical role in vehicle-to-pedestrian communication~\cite{Dey2017}, and is most effective when aligned and working in tandem with eHMI messages~\cite{Dey2020}. Thus, AV's movement has the potential to serve as an additional channel for embedding uncertainty information alongside eHMIs.

%Similarly, pedestrians typically rely on implicit cues, such as an approaching vehicle’s movement, to assess its intentions~\cite{Dey2017}, yet how these cues can effectively communicate AV uncertainty remains unclear.
In this study, we investigate the communication of an AV's uncertainty regarding its yielding decision to pedestrians in a typical unsignalised crossing scenario. This investigation is guided by the following research questions:

\begin{itemize}
    \item \textbf{RQ1}: How can the uncertainty of the AV system be effectively communicated to pedestrians?
    \item \textbf{RQ2}: What are the effects of uncertainty communication on pedestrians's perceived safety, trust and user experience?
\end{itemize}

%I was trying to address reviewers' comments on the framing of two concepts as implicit or explicit, as the reviewer mentioned : " explicit versus implicit cues but conflates this with display modality (screen vs. whole-vehicle movement). Implicit uncertainty cues could also be presented on-screen (e.g., using color [a]), meaning the comparison tests cue explicitness and display modality. The underlying design rationale should be further clarified or corrected." So I frame two concepts as eHMIs based Vs Motion based (in conjunction with eHMI showing intent)  Please check if my changes make sense. 
To address these questions, we investigate two approaches to communicating AV uncertainty: (1) an eHMI-based approach that displays the AV’s confidence level alongside its displayed yielding intent; and (2) a movement-based approach that embeds uncertainty information in the vehicle’s motion, working in tandem with an eHMI that displays the AV’s intent.
%To address these questions, we investigated both explicit (eHMI-based) and implicit (movement-based) approaches to uncertainty communication.
%we designed two uncertainty communication approaches (i.e., explicit and implicit) and .
We tested them in a crossing scenario where the AV's confidence level in detecting and predicting a pedestrian's crossing intent was either high or low. In a 2x2 within-subject virtual reality (VR) experiment (N=26), participants assessed the effectiveness of communication designs, perceived safety, trust and user experience through questionnaires. Our findings show that the explicit approach is more effective and preferred in communicating uncertainty to pedestrians, and that displaying a higher confidence level contributes to more positive perceptions of pedestrians to AVs.

This paper contributes: (1) the first empirical investigation into how AV uncertainty can be communicated to pedestrians, expanding the design space of eHMIs beyond intent signalling; and (2) evidence that explicit uncertainty cues, such as confidence displays, can influence how pedestrians interpret AV behaviour and their subjective experience.% and make crossing decisions.}

\section{Related Work}

\subsection{Uncertainty in Autonomous System}

The decision-making processes of autonomous systems inherently involve uncertainty, which can arise from sources such as noise in data input and imperfections in machine learning models~\cite{kochenderfer2015Uncertainty,Bhatt2021}.
Making such uncertainty transparent is essential to avoid over-reliance on system outputs~\cite{Bhatt2021,Prabhudesai2023}, particularly in high-stakes domains such as AI-assisted medical decision-making~\cite{begoli2019MedicalDecision} and public policy development~\cite{nordstrom2022ai}. Communicating uncertainty in autonomous systems has been shown to facilitate trust calibration between humans and automated systems~\cite{Bhatt2021,Helldin2013,Peintner2022}, fostering effective human-system collaboration~\cite{Schömbs2024,leusmann2023UncertaintyLoop}, and supporting adaptive autonomous decision-making~\cite{joslyn2013,Abraham2021AdaptiveAutonomy}. For instance, \citet{Prabhudesai2023} found that conveying uncertainty about machine learning predictions encourages more analytical thinking in users, thereby reducing over-reliance in AI-assisted decision-making.

While any AI system can ultimately influence real-world outcomes, embodied AI systems like AVs directly interact with their physical surroundings in real time, making errors in decision-making more consequential. In unpredictable urban traffic environments, uncertainty is further amplified by factors such as pedestrians’ ambiguous intentions and adverse weather conditions that can degrade perception capabilities~\cite{Yang2023Uncertainties}. 
Failing to communicate such uncertainty could potentially result in serious safety risks~\cite{Helldin2013}. 

%Consequently, extensive research has focused on communicating uncertainty to in-vehicle users to foster trust calibration and maintain situational awareness~\cite{doula2023Truth, Kunze2018,Kunze2018b,kunze2019,Helldin2013}. However, such information is largely concealed from vulnerable external stakeholders such as pedestrians who are directly affected by the AV’s decisions.

Importantly, uncertainty does not imply a malfunction. Rather, it reflects a functioning AV that communicates awareness of its own perceptual or predictive limitations. In in-vehicle contexts, these limitations are often communicated to drivers or passengers via internal HMIs, either prompting a takeover request or helping users maintain calibrated trust and situational awareness~\cite{doula2023Truth, Kunze2018,Kunze2018b,kunze2019,Helldin2013}. As Fridman noted in the context of shared perception~\cite{fridman2018humancentered}, the goal of such visualisations is not to present a flawless black-box system, but to help users understand what the vehicle can perceive and how confidently it makes decisions. For example, Tesla Full Self-Driving (FSD) and Waymo’s in-vehicle displays inform passengers whether an object has been recognised as a dog, cat, deer, or simply as an ambiguous white dot, allowing riders to better gauge the system’s limitations. 

This raises an important design question: Can such uncertainty information also be shared with external stakeholders? Specifically, could AV convey not only what it intends to do, but also how confident it is in making that decision to people outside the vehicles? The idea of mirroring information between internal and external HMIs has been previously proposed as part of holistic AV communication~\cite{dong2023holistic, dong2024exploring}, where internal displays might indicate that the AV is yielding, while the external display signals to the pedestrian that it is safe to cross~\cite{izquierdo2024pedestrian}. Extending this concept, our study explores the potential of externalising uncertainty, projecting the AV’s internal confidence outward, and investigates how such communication shapes pedestrian perceptions of safety, trust, and user experience.

\subsection{Uncertainty Communication Approaches}
%The definition from explicit communication (eHMI) and implicit communication (motion-dynamics and kinematics)~\cite{Dey2021ImplicitandExplicit}

An autonomous system's uncertainty can be communicated through either explicit or implicit means. Explicit uncertainty communication typically focuses on visualising probability~\cite{winkler2015}, with uncertainty often framed as the inverse of confidence. For example, in the context of autonomous driving, if the confidence level in identifying a road sign or a pedestrian's intention is low, the system's uncertainty is high and may necessitate human intervention~\cite{Peintner2022}. Approaches to explicitly present system confidence levels include displaying percentage numbers~\cite{Peintner2022} and graphics, such as bar graphs, icon arrays, and gauges~\cite{Peintner2022,Schömbs2024}. Some designs also utilise anthropomorphic features such as heart rate indicators to communicate real-time confidence levels~\cite{kunze2019}. %However, none of these explicit representations is designed for eHMIs, raising questions about the appropriate amount, distribution, and format of information for AV-pedestrian communication.
Implicit uncertainty communication, on the other hand, focuses on signalling hesitation, a commonly recognised indicator of uncertainty in human-human interactions~\cite{Schömbs2024}. Previous research on robots has examined cues for hesitation like pausing, slowing, delaying actions, and repeated pull-back movements as approaches to convey uncertainty~\cite{Schömbs2024,Yamada2013,Zhou2017}. %However, none of these implicit representations of uncertainty has been tested on AVs.

Although uncertainty communication can enhance transparency and support trust calibration~\cite{Bhatt2021,Prabhudesai2023}, communicating additional uncertainty information may increase people's cognitive demand and workload, consequently diminishing the efficiency of interactions~\cite{kunze2019,Peintner2022}. Interviews with human drivers revealed polarised attitudes regarding uncertainty displays~\cite{Peintner2022}, indicating interpersonal differences in the desired and accepted amount of information about the automation’s inner workings. Moreover, the level of confidence also impacts how humans perceive automation uncertainty~\cite{Hough2017}. For instance, communicating low confidence levels has been shown to reduce drivers’ trust towards the automated vehicle and negatively affect usability~\cite{Peintner2022}, indicating a sweet spot between fostering transparency and maintaining an appropriate level of confidence and usability in automated systems.
%Moreover, it is argued that people can perceive not only the presence of uncertainty but also its level~\cite{Hough2017}. Although this concept is further supported by the findings that compared with no uncertainty communication, displaying confidence level would negatively affect drivers’ trust and usability ratings of the AV~\cite{Peintner2022}, the effects of different confidence levels on human perception remain to be answered.

\subsection{AV-Pedestrian Communication}
The absence of human drivers in AVs eliminates established communication conventions between pedestrians and drivers~\cite{Schneemann2016driverGesture}, which have traditionally facilitated smooth co-navigation, especially in right-of-way negotiation. To address this gap, a variety of eHMIs have been developed to externalise an AV's status and intent to nearby pedestrians, communicating information such as (non-) yielding intent, 
situational awareness, and current or future maneuvers~\cite{Dey2020, FAAS2020INFO}. Communication methods include explicit approaches, such as lighting signals~\cite{Dey2020Color} or projected zebra crossings~\cite{Nguyen2019Porjection}, and implicit approaches, such as vehicle motions~\cite{Moore2019}. However, despite the development of numerous eHMIs in both research and industry, the inherent uncertainty of AVs decision-making remains unaddressed in communication with pedestrians. Moreover, studies indicate that the presence of eHMIs, especially those that simply display status information, can inadvertently foster overtrust among pedestrians~\cite{hollaender2019,faas2020longitudinal}. Communicating additional uncertainty information, such as the confidence level underlying an AV's decision, could more accurately reflect the vehicle's actual capabilities.

\section{Method}
To examine uncertainty communication in AV–pedestrian interactions, we visualise the AV’s confidence in detecting pedestrian intent and its yielding decision. Our VR study employed a 2×2 within-subject experiment design, manipulating two independent variables: communication approach~(explicit, implicit) and confidence level (high, low). The sequence of the conditions was counterbalanced for each participant using a balanced Latin Square design to diminish potential learning effects. Below, we describe the scenario, design concepts, and study setup in detail.

\subsection{Scenario}
The scenario took place on an unregulated two-lane street without traffic lights or pedestrian crossings, where pedestrians did not have right of way and had to interact with the AV to ensure safe crossing.  Night-time was chosen as a contextual factor to enhance the scenario's realism, making the uncertainty caused by reduced sensor visibility more believable for participants. Ambient urban audio (e.g., wind, distant traffic) was integrated into the virtual environment to enhance spatial presence. 

The simulated vehicle was SAE Level 5 and featured no visible driver. While most studies on AV-pedestrian interaction include both stopping and non-stopping behaviours to direct participants’ attention to vehicle kinematics or external displays~\cite{tran2021review}, our setup employed a consistent stopping behaviour. This allowed us to focus the investigation on how the AV communicates uncertainty during a stop, rather than confounding the experiment with varying yielding decisions. Moreover, our approach reflects a conservative decision-making strategy~\cite{Yang2023Uncertainties}, whereby the AV defaults to a safer fallback mode even when its confidence in a pedestrian’s crossing intent is low. Although the AV was set to yield in all conditions, participants were not informed of this in advance and were asked to assess safety before deciding whether to cross, mirroring real-world uncertainty and encouraging natural decision-making. The AV's initial speed was 30 km/h, in line with typical urban speed limits. It approached participants from their right-hand side, began braking at a distance of 35 m with the uncertainty design activated simultaneously, and ultimately stopped to yield.

\begin{figure*}[htbp]
    \centering
    \includegraphics[width=1\linewidth]{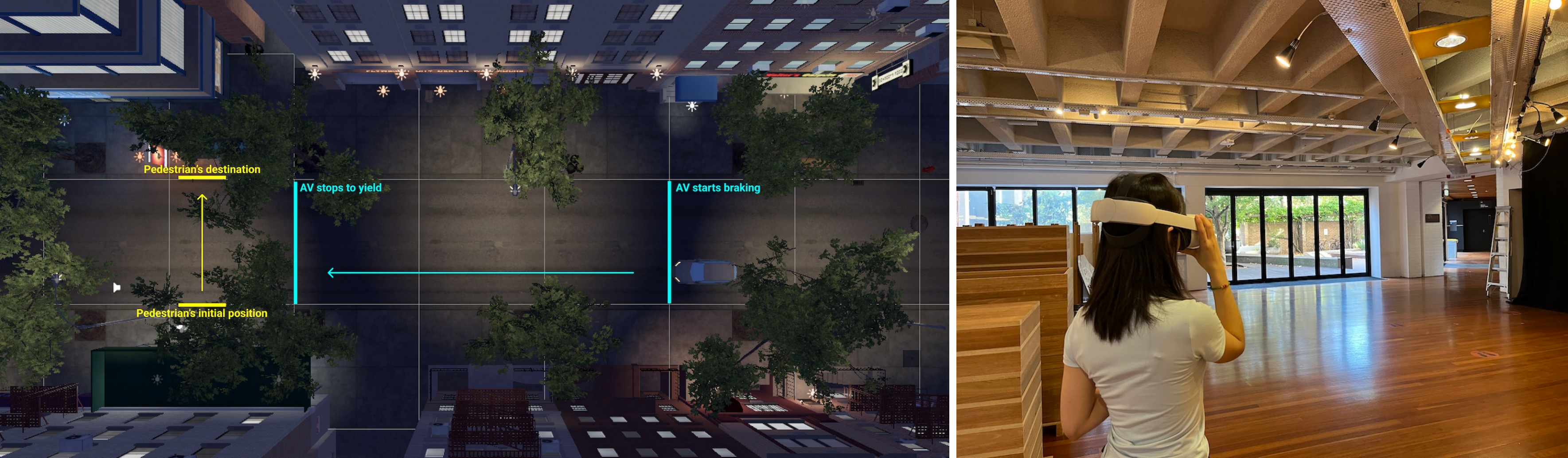}
      \caption{(a) Bird's-eye view of the tested scenario setting in VR. (b) On-site study environment for the VR experiment.}
      \Description{This figure displays a night-time urban setting from the bird's-eye view, illustrating the positions and movements in the experiment. The pedestrian is shown at their initial position on the left side of the road, with their destination marked on the opposite side. The autonomous vehicle (AV) is shown starting its braking procedure, positioned further down the road. The figure includes directional arrows to indicate the pedestrian's path and the AV’s stopping point. The dark environment highlights the key elements: the pedestrian's crossing task and the AV's approach and yield behaviour.}
    \label{fig:scenario}
\end{figure*}

Each trial began with participants positioned on one side of the road, and their task was to safely cross to reach the bus station on the opposite side (see~\autoref{fig:scenario}, a). Study participants wore a Meta Quest 2 headset to interact with the AV. The experiment was conducted in an open indoor space (6 m × 11 m), allowing participants to physically cross the street within a virtual environment rendered at a 1:1 scale (see~\autoref{fig:scenario}, b). 

\subsection{Uncertainty Communication Design}
\begin{figure*}[htbp]
    \centering
    \includegraphics[width=1\linewidth]{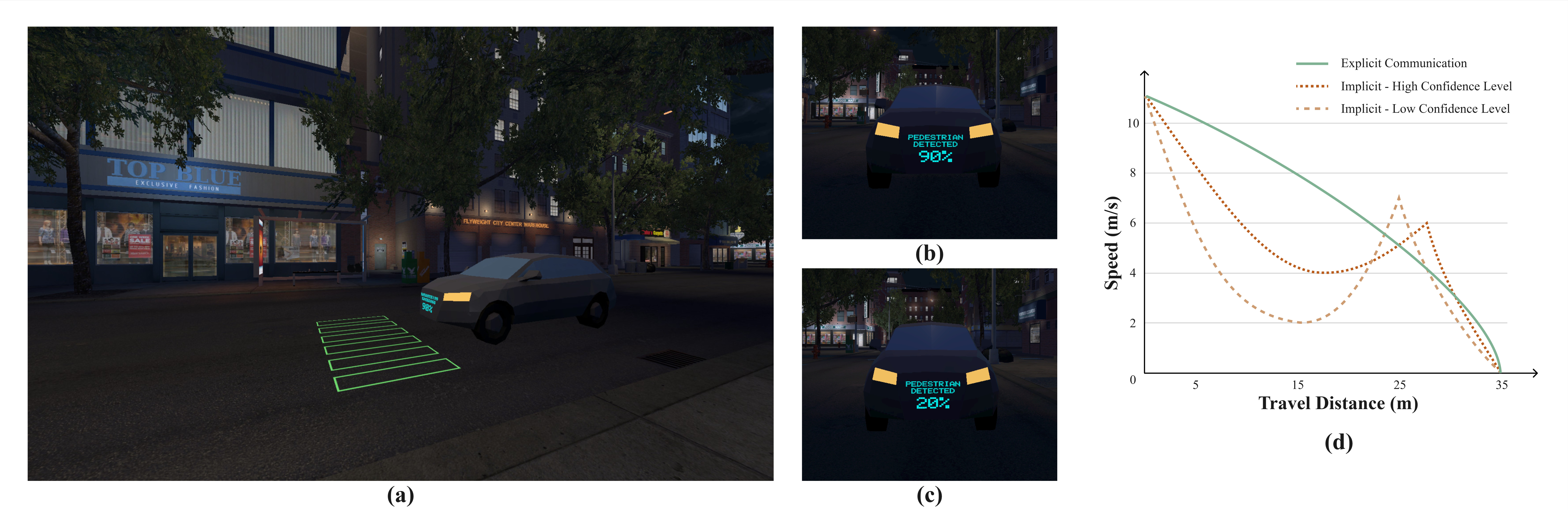}
      \caption{(a) AV communicating its intent to yield via green zebra crossing projection and explicitly displaying its confidence level. (b) eHMI for Explicit-High Confidence Level. (c) eHMI for Explicit-Low Confidence Level. (d) Implicit motion patterns for different conditions.}
      \Description{This figure consists of four subfigures. Subfigure (a) presents the scenario from the pedestrian’s first-person view at the unmarked crossing. In the foreground, the AV is seen on the right-hand side with a green zebra crossing projected onto the street. The AV’s front windshield displays "Pedestrian Detected 90\%". On the opposite side of the road, the bus stop and surrounding buildings are visible, providing context to the pedestrian’s position and destination across the street. Subfigure (b) and (c) both show the autonomous vehicle (AV) from the front, with the uncertainty message displayed on its windshield. In subfigure (b), the message reads "PEDESTRIAN DETECTED 90\%" in bright, pixelated cyan text. The AV’s headlights are illuminated in yellow, and the dark urban environment is visible in the background. Subfigure (c) is similar, except for the percentage is 20\%. SUbfigure (d) presents a graph showing the AV’s speed (in meters per second) against travel distance (in meters) for three different communication conditions: explicit communication (green curve), implicit high-confidence level (dotted orange curve), and implicit low-confidence level (dashed orange curve).The green curve (explicit communication) shows a smooth, continuous deceleration from 10 m/s to a stop, following a steady downward slope. The dotted orange curve (implicit high-confidence level) also shows a deceleration, but it is more gradual, followed by a slight acceleration (4 m/s to 6 m/s) before decelerating again. The dashed orange curve exhibits a more erratic pattern, with sharp fluctuations in speed, indicating abrupt deceleration (11 m/s to 2 m/s) and acceleration (2 m/s to 7 m/s) phases, reflecting higher uncertainty in the vehicle’s movement.}
    \label{fig: communication design}
\end{figure*}

We incorporated the green zebra crossing design from Mercedes-Benz’s F 015 projection concept~\cite{mercedes2015} across all conditions to indicate the AV’s yielding intent (see~\autoref{fig: communication design}, a). The projection was activated when the AV stopped to give way. This design was chosen due to its familiarity and the positive reception reported among participants in~\cite{Locken2019}. On top of that, we developed approaches to communicate the uncertainty associated with this yielding intent. We designed both explicit and implicit uncertainty communication concepts to leverage their respective strengths in communication effectiveness.% and minimising information overload.%, in order to explore their application in uncertainty communication.

%, integrating both explicit and implicit approaches to leverage their respective strengths. 
%Explicit communication compensates for the absence of typical human communication cues, such as eye contact and gestures, commonly used by human drivers~\cite{ACKERMANN2019,Habibovic}. In contrast, implicit communication relies on behaviours that serve a functional purpose while simultaneously enabling observers to intuitively infer the vehicle's state or intent~\cite{Dey2017}.

%Through informal testing within the research group, we found the confidence percentage clearer to perceive and less misleading from a pedestrian’s perspective. 

%After internal discussion, we selected the percentage format, as it was considered clearer to to interpret and less prone to misinterpretation from a pedestrian’s perspective.

Based on previous work, visuals are the most commonly used modality for explicit uncertainty communication~\cite{Schömbs2024,kunze2019,Peintner2022}. We compared two visual designs from previous proposed for communicating uncertainty to drivers of automated vehicles: a confidence bar and a confidence percentage~\cite{Peintner2022}. In the context of pedestrian-AV interaction, crossing decisions are typically made quickly, requiring signals that are both clear and easy to interpret. Through internal discussion, we determined that a confidence percentage was better suited to this setting than a graphical bar. While bars are commonly used in visual displays, they can be ambiguous without explicit reference points. In contrast, percentages offer a more direct and familiar way to express system confidence.

The final design featured a `\textit{Pedestrian Detected}' message and confidence percentage displayed on an LED screen at the front of the vehicle—an optimal position for visibility~\cite{zheng2024exploring} (see~\autoref{fig: communication design}, b, c). The texts were displayed in cyan—a colour considered neutral for eHMIs~\cite{Dey2020Color}. The confidence percentages were `\textit{90\%}' for the high level and `\textit{20\%}' for the low level~\cite{mccaffery2012}. Although these percentages do not represent actual model outputs, they serve as designed proxies for high and low confidence. We selected these values to create a noticeable contrast while avoiding extreme figures (e.g., 100\% or 0\%) that could imply false certainty or failure, thereby supporting clear yet realistic interpretation of the AV’s internal state.

%The uneven motion pattern was inspired by a pausing motion pattern previously designed for robotic arms to signal hesitation~\cite{Schömbs2024, Zhou_2017}. To ensure smooth speed transitions, we utilised Unity's linear interpolation function to manipulate the speed change in implicit conditions. 

For the implicit communication approach, we designed a three-stage motion pattern to encode confidence levels. The AV's braking-to-yielding behaviour was divided into three phases: initial deceleration, acceleration, and final deceleration to a stop. This contrasts with the stable and continuous fixed rate of deceleration at 1.7 m/s² used by the AV in the explicit conditions. The irregular motion was inspired by previous work on robotic arms, where pausing or non-linear movements were used to signal hesitation or uncertainty~\cite{Schömbs2024, Zhou_2017}. 

To create this effect, we used Unity’s linear interpolation function to control the AV’s speed over time in a smooth but deliberately uneven way. In the high-confidence condition, the AV decelerated from 11 m/s to 4 m/s, then accelerated gently to 6 m/s. In contrast, the low-confidence condition featured a more erratic motion: the AV decelerated sharply from 11 m/s to 2 m/s during the initial phase, then accelerated more abruptly to 7 m/s before stopping (see~\autoref{fig: communication design}, d).

%compared to the high-confidence condition, with more abrupt speed changes to indicate a higher level of uncertainty. Specifically, the AV in the low-confidence condition 
%corresponding to the "90\%" confidence level in the explicit approach%To ensure smooth speed transitions, we utilised Unity's linear interpolation function. %For high confidence levels, the AV decelerated from 11 m/s to 4 m/s, accelerated from 4 m/s to 6 m/s, and then decelerated from 6 m/s to a stop, creating a relatively stable deceleration profile. For low confidence levels, the AV decelerated from 11 m/s to 2 m/s, accelerated from 2 m/s to 7 m/s, and decelerated from 7 m/s to a stop, producing a more erratic and unstable motion pattern. 
%The high-confidence condition presented a slightly uneven pattern, corresponding to the "90\%" confidence level in the explicit approach, while the low-confidence condition featured a more erratic motion pattern, indicating greater uncertainty. This design was %However, we further refined the motion pattern to ensure a smooth and natural flow, avoiding interpretations of malfunction by pedestrians. 

\subsection{Procedures}
A total of 26 participants (15 women, 11 men) between the ages of 18 to 34 (\textit{M} = 25.04, \textit{SD} = 2.14) participated in our study. We recruited the participants through our university's notice boards and social media. This study received ethical approval from the University of Sydney Human Research Ethics Committee (HREC), protocol 2023/HE000434.

Each session began with a brief introduction to the study, followed by the participants filling out a consent form and demographic questionnaire. Participants then were provided with instructions on how to wear and operate the VR equipment. Prior to the experiment, each participant went through a familiarisation session to practise crossing the street and returning to their initial position. During this session, an AV approached from the right-hand side without decelerating or yielding. Participants then proceeded to experience all four experimental conditions. Participants were not provided with explanations of the communication concepts prior to the experiment.
After each condition, participants removed the VR headset and completed the questionnaires.
Upon completing all conditions, participants participated in a 10-minute semi-structured interview. Each session lasted approximately 60 minutes.

%, alongside three standardised questionnaires to assess perceived safety, trust, and user experience. %The three original questions are developed to assess the visibility, confidence, and intuitiveness of the design. 
%Three custom-designed questions were used to evaluate the visibility, confidence, and intuitiveness of the design. 

\subsection{Measures}

To evaluate how effectively the designs conveyed AV uncertainty, we focused on three key aspects: visibility, clarity of the conveyed AV confidence, and intuitiveness. These measures were chosen to capture both the perceptual salience and interpretability of the communication approaches, which are essential for real-time pedestrian decision-making. Three custom-designed questions assessed these aspects: 

\begin{itemize}
\item \textit{Visibility}: For the explicit communication approach, the visibility question asked, `\textit{How clearly did you notice the information displayed on the vehicle?'}. For the implicit communication approach, the visibility question was phrased as \textit{`To what extent did you notice anything special about the vehicle’s deceleration?'} to align with implicit communication through motion. Both used a 7-point Likert scale from `\textit{Not at all}' to `\textit{Very clearly}.' 

\item \textit{Confidence and Intuitiveness}: The confidence and intuitiveness item remained the same across all conditions, with the confidence question phrased as, `\textit{The information suggested that the vehicle was...}' rated from `\textit{Very uncertain}' to `\textit{Very certain},' and the intuitiveness question phrased as, `\textit{The information was...}', rated from `\textit{Not intuitive at all}' to `\textit{Very intuitive}.' 
\end{itemize}

In addition to evaluating the interpretability of the designs themselves, we also sought to understand how these uncertainty communication approaches shaped participants’ overall interaction experience, using constructs commonly used in AV-pedestrian research. All questionnaire items were rated on a 7-point Likert scale.

\begin{itemize}

\item \textit{Perceived Safety}: We adapted questions used in  ~\cite{Locken2019}, which asked: `\textit{The communication offers safety},' `\textit{The vehicle’s signals can be clearly perceived},' and `\textit{I perceived crossing the street as risky}.' 

\item \textit{Trust}: Trust was measured using the Trust in Automation Scale~\cite{jian2000}.

\item \textit{User Experience}: User experience was measured using the User Experience Questionnaire (UEQ)~\cite{laugwitz2008}. The UEQ includes six dimensions, attractiveness, perspicuity, efficiency, dependability, stimulation, and novelty.

\end{itemize}

%The confidence and intuitiveness item remained the same across all conditions.%The questions were phrased slightly differently to be in line with the uncertainty communication in each specific condition.

%Three questionnaires are used to assess perceived safety, trust, and user experience. 
%All questionnaire items were rated on a 7-point Likert scale. %The UEQ includes six dimensions, attractiveness, perspicuity, efficiency, dependability, stimulation, and novelty, which were examined in the following data analysis.

% List: https://www.overleaf.com/learn/latex/Lists
% Table: https://www.overleaf.com/learn/latex/Tables

\subsection{Data Analysis}
Repeated-measures ANOVA tests were conducted in SPSS to examine the main effects of the two variables: communication approach and confidence level. The interaction effect between communication approach and confidence level was also examined to assess their combined influence. The interview recordings were transcribed and analysed by the first author using an inductive thematic approach~\cite{braun2006thematic} to uncover patterns and themes.

\section{Results}

\subsection{Visibility, Confidence, and Intuitiveness}
\begin{figure*}[h]
\begin{center}
\includegraphics[width=0.8\textwidth]{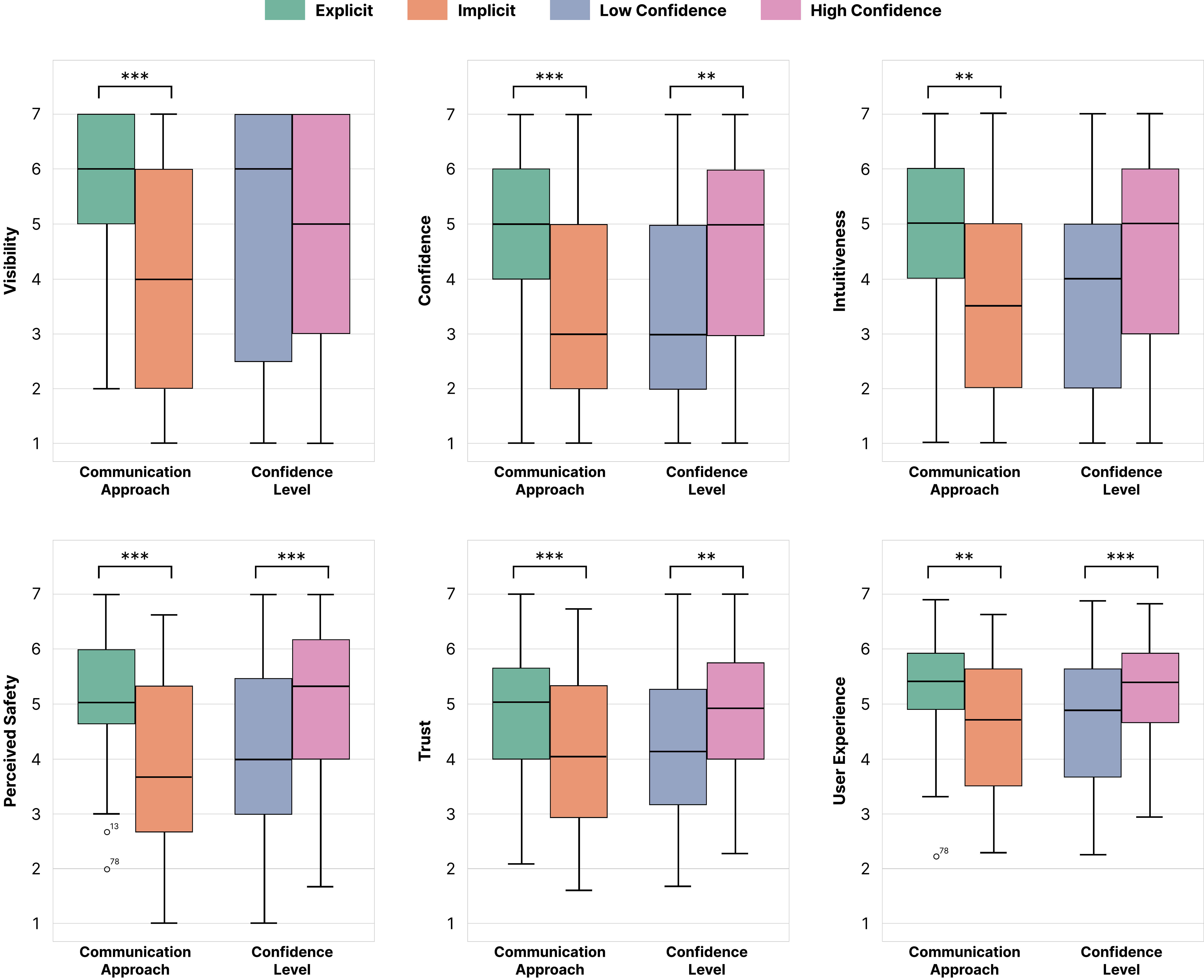}
\end{center}
\caption{Box plots showing the effect of communication approach (explicit vs. implicit) and confidence level (low vs. high) on six dependent variables: Visibility, Confidence, Intuitiveness, Perceived Safety, Trust, and User Experience.}\label{BoxPlots}
\label{fig:BoxPlots}
\Description{}
\end{figure*}

Descriptive analysis showed that explicit uncertainty communication was rated as more visible and intuitive, and made the AV appear more confident compared to implicit communication via movements (see~\autoref{fig:BoxPlots}). The ANOVA results revealed significant main effects of communication approach on visibility (\textit{F}(1, 25) = 25.66, \textit{p} < .001, $\eta^2_p$ = .51), confidence (\textit{F}(1, 25) = 24.00, \textit{p} < .001, $\eta^2_p$ = .49), and intuitiveness (\textit{F}(1, 25) = 11.17, \textit{p} = .003, $\eta^2_p$ = .31). Results also showed that confidence level had a significant main effect on the perceived confidence of the AV (\textit{F}(1, 25) = 9.94, \textit{p} = .004, $\eta^2_p$ = .28), with a significantly higher rating in high confidence level conditions than in low confidence level conditions (see~\autoref{fig:BoxPlots}). The results showed no significant main effects of confidence level on visibility (\textit{F}(1, 25) = 0.42, \textit{p} = .522) and intuitiveness (\textit{F}(1, 25) = 4.00, \textit{p} = .056). 

\begin{figure*}[h]
\begin{center}
\includegraphics[width=0.8\textwidth]{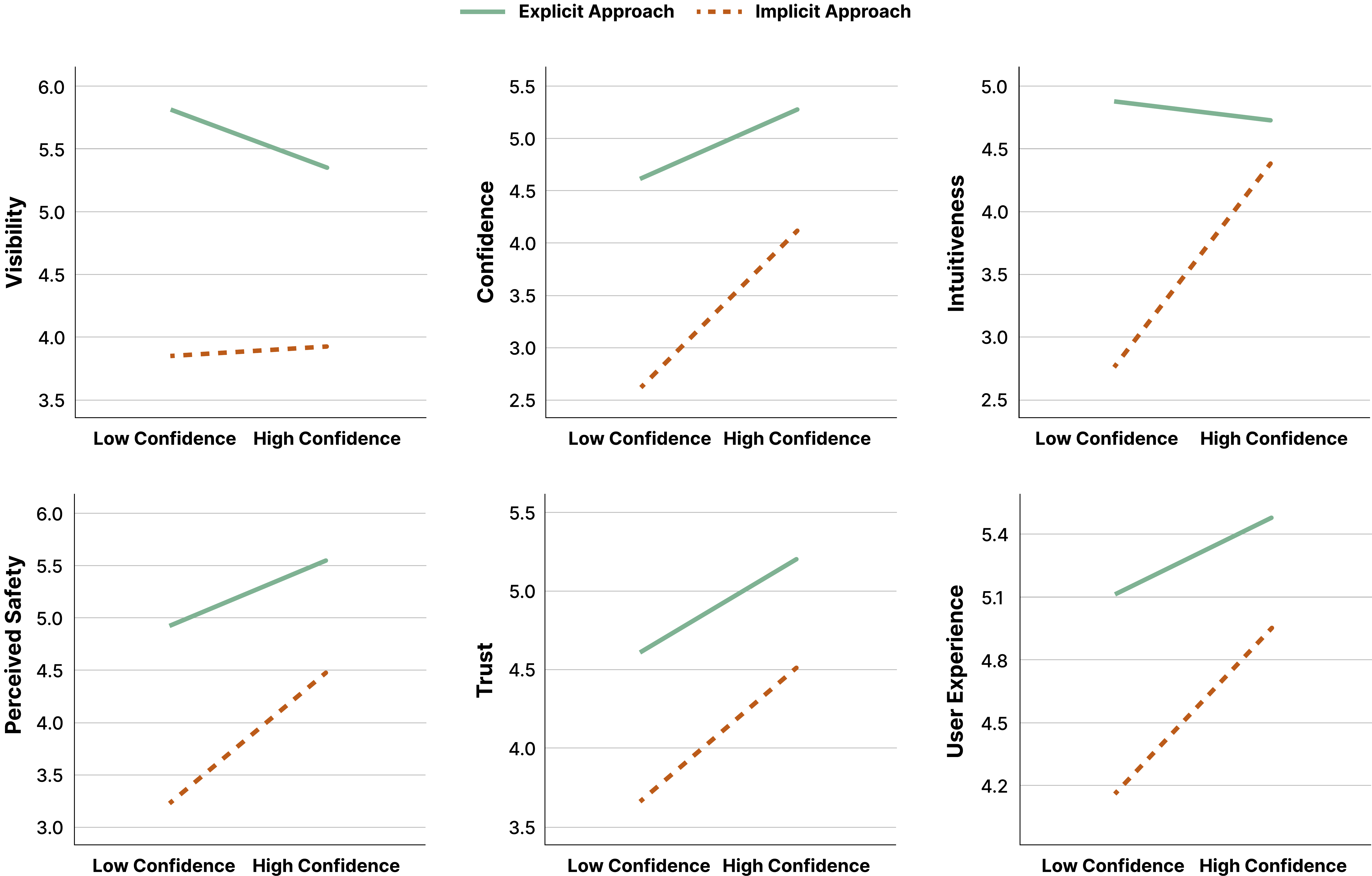}
\end{center}
\caption{Interaction between communication approach and confidence level across six measures. Each line represents the estimated marginal means for each measure.}\label{fig:interaction_plot}
\label{fig:box plots}
\Description{This figure presents interaction plots showing how the communication approach (explicit vs. implicit) and confidence level (low vs. high) interact across six measures: Visibility, Confidence, Intuitiveness, Perceived Safety, Trust, and User Experience. The y-axis represents the estimated marginal means for each measure, and the x-axis shows the confidence levels (low vs. high). The green line corresponds to the explicit communication approach, while the orange line represents the implicit approach. This figure highlights the differential responses in different conditions.}
\end{figure*}

For intuitiveness, there was an interaction effect between communication approach and confidence level (\textit{F}(1, 25) = 12.07, \textit{p}= .002, $\eta^2_p$ = .33). In implicit conditions, low confidence was perceived as significantly less intuitive than high confidence, whereas explicit conditions showed no significant difference between confidence levels (see~\autoref{fig:interaction_plot}). Interestingly, a descriptive assessment of the interaction plots also suggests that, on average, participants perceived the vehicle as more confident when using the explicit communication approach, even at low confidence levels, compared to implicit communication. No interaction effect was found on visibility and perceived confidence.

\textit{Qualitative feedback}: %The implicit communication approach with low confidence level condition scored the lowest in all of the three dimensions, which was also evidenced by the interviews showing participants’ varied understanding of the uncertainty communication. 
%It did not become clear, if participants were told what the numbers on the eHMI meant, as in the qualitative results in 4.1, more than half the participants apparently thought it was something else than confidence. Yet this is never discussed. This influences all ratings that participants did and their feedback, it means that the presentation of the information might not be suited. Yet authors claim overwhelmingly that the approach is working. This needs to be properly and critically discussed, which is completely missing. This is my main issue with the paper in its current form. Authors might be able to fix this in the short time provided for LBW but I think it is too much. 
%The finding that explicit uncertainty communication was rated higher in visibility is supported by participant's reflections during interviews. 
During the interviews, explicit text-based uncertainty communication was described as more clearly visible compared to vehicle motion, particularly under conditions with limited visibility. For example, P3 noted, \textit{`At night, it’s hard to notice if a vehicle is decelerating or accelerating, [while as] the text provides clarity.'} Similarly, P1 described the text display as \textit{`bright and relatively large.'} However, some participants commented on limitations related to the spatial positioning of the display. As the vehicle approached from the side, the front-mounted display—initially facing the pedestrian—became harder to read due to the increasingly oblique viewing angle. P12 remarked, \textit{`The display’s position could be improved. It’s on the car’s side, making it harder to see.'}

When asked in the post-study interviews about their interpretation of the two communication approaches, participants’ responses provided additional insight into the perceived intuitiveness of each. Explicit communication was most commonly understood as confidence or certainty (n=7) and accuracy in detecting the pedestrian's presence (n=6). However, despite the use of a text-based display, some participants misinterpreted the numeric value as representing the vehicle's battery level (n=4). P15 mentioned that they \textit{`at first thought [the display] was an ad'}, indicating potential issues with the contextual clarity of text-based eHMI displays. For implicit communication, 7 participants mentioned expressing uncertainty or hesitation in detection (n=7) when describing the vehicle's motion. Others interpreted the motion as ambiguous or malfunctioning system behaviour, for example, system errors (n=3), AV’s changing its mind (n=3), and confusion (n=2).

\subsection{Perceived Safety}
Descriptive analysis of the box plots (see~\autoref{fig:BoxPlots}) suggests that participants felt safer when uncertainty was communicated explicitly and when the AV displayed a high confidence level. Furthermore, the interaction plots show that participants felt safest in the condition where both of these factors were combined, that is, when uncertainty was communicated explicitly and the confidence level was high (see~\autoref{fig:interaction_plot}). The ANOVA results showed significant main effects of both communication approach and confidence level on perceived safety (\textit{F}(1, 25) = 21.44, \textit{p} < .001, $\eta^2_p$ = .46; \textit{F}(1, 25) = 19.80, \textit{p} < .001, $\eta^2_p$ = .44). No interaction effect between communication approach and confidence level was found on perceived safety.

\textit{Qualitative feedback}: In the interview, 12 participants reported that the uncertainty communication made them feel safer because this gave them more information (n=5) and more control (n=2). For example, P17 stated, \textit{`it helps me make better decisions [...] if the car shows 90\%, I’d feel safe to cross, but if it’s 20\%, I’d wait'}, and P26, \textit{`if the car tells me it’s uncertain, I can adjust my actions accordingly.'} Conversely, 10 participants reported decreased perceived safety, some of them explicitly attributing it to heightened awareness of potential danger (n=3). Four participants stated that they would only feel safer if the displayed confidence level was high. For example, some participants mentioned that they wouldn't initiate crossing until the vehicle's confidence level hits 100\%, with P2 stating, \textit{`even 95\% or 85\% confidence feels insecure [...] when crossing the street'}.

\subsection{Trust}
Both communication approach and confidence level had significant main effects on trust (\textit{F}(1, 25) = 13.90, \textit{p} = .001, $\eta^2_p$ = .36; \textit{F}(1, 25) = 11.00, \textit{p} = .003, $\eta^2_p$ = .31), which indicated that participants trusted the AV more when a high confidence level was explicitly displayed (see~\autoref{fig:BoxPlots}). No interaction effect between communication approach and confidence level was found on trust.

\textit{Qualitative feedback}: In the interviews, 12 participants indicated they would trust an AV more if it communicated its uncertainty, noting that \emph{`the communication suggests the system’s honesty'}~(P20). On the other hand, eight participants indicated a decrease in trust as they doubted that an AV should make mistakes (n=2). Four participants reported that their trust would only increase if the displayed confidence level was high. Interview responses further highlighted the challenge of designing unambiguous messages for text-based eHMIs. While the presence of text (i.e., `Pedestrian Detected') was associated with increased trust---particularly when interpreted as a status message---the accompanying display of low confidence percentages prompted participants to question the vehicle’s decision-making. As P20 noted: \textit{`The text itself increases trust, but the confidence percentages (like 20\%) make me question the vehicle’s actions.}'

\subsection{User Experience}
The ANOVA results revealed significant main effects of both communication approach and confidence level on user experience (\textit{F}(1, 25) = 10.89, \textit{p} = .003, $\eta^2_p$ = .30; \textit{F}(1, 25) = 16.97, \textit{p} < .001, $\eta^2_p$ = .40). Participants rated user experience higher when the AV’s uncertainty communication was explicit and confidence level was high (see~\autoref{fig:BoxPlots}). No interaction effect between communication approach and confidence level was found on user experience.

\textit{Qualitative feedback}:  
Qualitative data showed that most of the participants preferred the explicit approach. Participants in favour of the explicit approach (n=19) reported that explicit communication was clearer, more straightforward (n=6) and efficient (n=3) than implicit approach, which some viewed as ambiguous (n=4), hard to notice (n=3), and dangerous (n=3). Participants who preferred the implicit approach (n=7) thought it was more noticeable (n=2), more intuitive (P2), and more accessible (P16). Fifteen participants believed that it is beneficial to communicate uncertainty because pedestrians have the right to know (n=4), and the output of AI systems needs to be reviewed by humans (n=2). P22 commented: \emph{`This information feels transparent, and the AV should not be hiding things from pedestrians.'} In contrast, eight participants thought it was unnecessary to communicate uncertainty, expressing concerns about potentially inducing negative reactions from pedestrians (n=4) and overwhelming them with excessive information (n=2). As P10 stated, \emph{`Why do I need this much information about a car when I'm just doing a simple thing like crossing? All I need to know is whether I can cross.'} Contrary, P14 stated that \textit{`[they] think it's necessary [to express uncertainty] because the technology is still new'}, suggesting that transparency about system limitations may be important for calibrating trust during the initial rollout of AVs, but could become less critical as the technology matures and user familiarity increases.

\section{Discussion}

%%I think the qual results also touch on the aspect that pedestrians want 100% certainty or at least are expecting a specific treshhold. This paper touches on this in a different decision-making context: https://dl.acm.org/doi/10.1145/3581641.3584033 They argue that people move from heuristic-based reasoning to analytical reasoning.

\subsection{Comparing Explicit and Implicit Uncertainty Communication}

Our study found that explicit uncertainty communication was more visible and intuitive than implicit communication, particularly at low confidence levels. We observed a clear interaction between communication style and confidence: while participants consistently understood explicit messages, implicit cues became harder to interpret when confidence was low, likely due to irregular deceleration patterns that led to misinterpretation. This aligns with findings by \citet{schmidt2019hacking}, who showed that pedestrians interpret vehicle kinematics through a social lens, often perceiving unexpected motion as reactive, erratic, or even hostile. Deviations from familiar movement conventions appear to undermine interpretability. While vehicle kinematics are a crucial cue for pedestrian crossing decisions~\cite{Moore2019, Mahadevan2018}, these results highlight the need for caution when deliberately manipulating motion to convey additional information.

Participants reported higher trust in the AV, higher perceived safety, and greater overall user experience when uncertainty was communicated explicitly. These findings contrast with \citet{Schömbs2024}, who observed consistently high trust towards a robot in a collaborative decision-making task, regardless of whether communication was explicit or implicit, and independent of confidence levels. A possible explanation lies in the differing contexts: AV-pedestrian interactions involve safety-critical decisions, whereas engaging with a robot arm presents a lower-stakes scenario. This suggests that the effectiveness of uncertainty communication may depend not only on the communication modality but also on the perceived risk associated with the interaction.

Interestingly, AVs that explicitly communicated uncertainty were perceived as more confident overall, even when conveying low confidence. This may be the result of the AV exhibiting a more stable and natural deceleration in the explicit conditions compared to the implicit ones. This finding indicates the significant role of AV motion in shaping pedestrians' judgements of AVs, aligning with \cite{Dey2017}, which suggested that pedestrians primarily rely on implicit cues like movement patterns to interpret AV intentions. Similarly, \citet{Schömbs2024} found that robot behaviour was more intelligible at high confidence, suggesting that uncertainty conveyed through motion alone is less effective.

In summary, our findings suggest that implicit uncertainty communication is less intuitive for pedestrians, particularly at low confidence levels, but pedestrians may still rely on it to make judgements. That said, future work could explore a hybrid approach—combining explicit and implicit displays—to mitigate each approach's limitations. For instance, an explicit+implicit display may enhance accessibility for groups such as children.

\subsection{When and Why Uncertainty Communication Matters}

In our design, implicit uncertainty communication conveyed hesitation through AV motion, while explicit communication displayed the precise values of AV's confidence levels, giving users insight into its decision-making process. Providing users with access to data within the autonomous decision-making process can significantly enhance system transparency, fostering trust and improving interaction quality~\cite{schmidt2020transparency}. For instance, \citet{Arshad2015} revealed that conveying uncertainty with known probabilities enhances user confidence, while ambiguity from unknown probabilities reduces it. Though their study focused on expert users who possess domain-specific knowledge of AI, our findings suggest that pedestrians—without specialised knowledge—also prefer transparent information on system uncertainty.

An additional consideration is the potential overtrust issue in eHMIs~\cite{faas2020longitudinal}. Overtrust in AVs can lead to pedestrians relying too heavily on system signals without critically assessing the traffic situation. \citet{faas2021calibrating} addressed trust calibration by designing an eHMI system that distinguished between two functions: a status eHMI indicating automated driving mode and an intent eHMI signalling the AV’s intent to yield. Their approach provided transparency by ensuring that if the yielding intent signal did not activate, it indicated a malfunction or misjudgment. They suggested that pedestrians should be educated about AV misjudgments and how to detect potential malfunctions through eHMI cues. Similarly, several eHMIs have been designed to adapt their communication based on the AV’s confidence in interpreting the surrounding traffic; for example, projecting a green crosswalk when the vehicle is confident it is safe to cross, and a red crosswalk when pedestrians should proceed with caution~\cite{tran2024exploring}. Our study extends this direction by focusing on a different aspect of transparency, not just a binary function/malfunction distinction, but the degree of confidence in AV decision-making. Rather than simply indicating whether an AV will stop, our findings highlight the importance of conveying nuanced levels of confidence, ensuring that pedestrians are not only informed of AV intent but also its reliability in assessing the situation.

A recent study by \citet{cumbal2025visualising} investigated uncertainty communication in the context of delivery robots. Conducted in a lower-stakes domain, their work found that confidence communication had no significant impact on most measured variables, aside from a positive effect on predictability. Participants also expressed mixed views about the usefulness of confidence displays. In contrast, our findings demonstrate that explicitly displaying AV confidence can significantly enhance perceived safety, trust, and user experience, particularly when the AV exhibits high confidence. This highlights the importance of context: in high-risk environments such as road crossings, clear communication of system uncertainty may play a more critical role in shaping pedestrian decision-making and trust.

While both our study and that of \citet{cumbal2025visualising} focus on direct interaction, where uncertainty communication supports coordination between the robot and the human user, \citet{yu20205peek} explores a different form of engagement. Their work examines how delivery robots might express internal uncertainty to bystanders in public urban environments, not to negotiate shared decisions, but to subtly prompt voluntary assistance. This contrast highlights the varying roles of humans in uncertainty communication: from passive observers to active participants in safety-critical interactions.

\subsection{Limitations and Future Work}

Our study presents the first investigation into AV uncertainty communication to pedestrians. While offering valuable insights, several limitations should be acknowledged. 

First, the findings are based on a relatively small sample comprising mostly university students and young professionals. The AV uncertainty communication was evaluated in a controlled VR setting, which, while providing experimental consistency, does not fully replicate real-world environmental complexities. In addition, the uncertainty communication concepts were paired with a projected green zebra crossing design, making it unclear how effective they would be when integrated with other eHMI concepts. %Future research could explore the effectiveness of uncertainty communication when combined with alternative eHMI designs.

Another limitation concerns the static nature of our explicit communication design. The confidence percentage was shown at the onset of the AV's deceleration, remaining fixed throughout the interaction. In real-world scenarios, however, an AV’s confidence in a pedestrian’s crossing intent is likely to evolve over time as it observes behavioural cues such as motion, hesitation, or gaze direction. Future work could explore dynamic or updating confidence displays that reflect these real-time fluctuations, and examine how such updates influence pedestrian interpretation, trust calibration, and decision-making. 

We also acknowledge that our explicit communication approach assumes the presence of a dedicated display on the AV, which may not always be feasible in practice. Constraints such as limited visibility, information overload, or heterogeneous vehicle designs could make physical displays less reliable. In such cases, augmented reality (AR) offers a promising alternative for conveying uncertainty information in more flexible and context-sensitive ways. Future AR systems, such as wearable AR glasses~\cite{tabone2021towards} could allow for personalised overlays that adapt to pedestrian needs without relying on physical vehicle, mounted displays. This direction is particularly relevant during the initial deployment of AVs, when additional communication channels may be needed to establish public understanding and trust. Exploring AR as a platform for uncertainty communication presents a compelling avenue for future work.

Finally, while our study examined how uncertainty communication influences pedestrian trust and decision-making, long-term effects remain unexplored. Future research should investigate whether repeated exposure to uncertainty communication improves pedestrian trust calibration over time or leads to unintended behavioural changes.

%Despite the meaningful findings across multiple variables, all the data collected in this study are subjective. Future research could incorporate objective measures, such as crossing behavior and decision-making time, to gain a more comprehensive understanding of the outcomes of uncertainty communication. 

%Additionally, as all scenarios in this study were conducted at night-time, the usability of the eHMI design during daytime conditions remains uncertain. Future studies could explore eHMI designs optimized for daytime use and investigate strategies to enhance the usability, accessibility, and scalability of uncertainty communication across diverse contexts.

\section{Conclusion}
%We conducted a VR study to investigate the effects of communicating AV uncertainty to pedestrians. 

This paper offers the first exploration into how AVs might communicate their internal uncertainty to pedestrians, a dimension absent from current eHMI design. Our findings show that explicit uncertainty communication is easier for pedestrians to interpret, particularly when AV confidence is low. It also led to higher ratings of trust, perceived safety, and overall user experience, regardless of the actual confidence level conveyed. These results demonstrate the importance of transparent and interpretable uncertainty cues in AV-pedestrian interaction, especially during early deployments when public trust must be earned. 

While exploratory in scope, our findings highlight the potential of treating uncertainty as a communicative resource in AV design. We hope this work encourages further investigation into how AVs can transparently express the reliability of their intent, and how future interfaces, whether through visual signals, vehicle motion, or AR, can adapt to communicate uncertainty in more effective and context-aware ways.

%The findings indicate that explicit uncertainty communication is easier for pedestrians to understand, particularly when the AV's confidence level is low. In addition, participants reported higher trust, perceived safety, and overall user experience when uncertainty was communicated explicitly, regardless of AV's confidence level. 
%At the same time, our study surfaces potential challenges in conveying internal system information in ways that are interpretable, contextually meaningful, and cognitively manageable for pedestrians.

%This study provides valuable insights for designing eHMIs to communicate AV uncertainty, contributing to the growing body of knowledge in this emerging field. 

%It also identifies potential directions for future research to further enhance the effectiveness of eHMIs in pedestrian-AV interactions.

\begin{acks}
We sincerely thank the capstone research unit coordinator Joel Fredericks for his guidance and support, the participants for their valuable contributions, and the anonymous reviewers for their insightful feedback. This research is supported by the Australian Research Council (ARC) Discovery Project DP220102019, Shared-space interactions between people and autonomous vehicles.
\end{acks}

\bibliographystyle{ACM-Reference-Format}
\bibliography{references}

\appendix

\end{document}